\documentclass[letterpaper,11pt]{article}
\usepackage{graphicx}
\usepackage{comment}
\usepackage{url}

\begin{document}

\newcommand{\singlespace}{\renewcommand{\baselinestretch}{1} \normalsize}
\newcommand{\doublespace}{\renewcommand{\baselinestretch}{1.1} \normalsize}

\title{Reducing the Bias and Uncertainty of Free Energy Estimates by Using
Regression to Fit Thermodynamic Integration Data}
\author{Conrad Shyu\\
\url{conrads@uidaho.edu}\\
 \\
Department of Physics\\
University of Idaho\\
Moscow, ID 83844-0903\\
\and
F. Marty Ytreberg\\
\url{ytreberg@uidaho.edu}\\
 \\
Department of Physics\\
University of Idaho\\
Moscow, ID 83844-0903}

\maketitle \doublespace

%
\begin{abstract}
This report presents the application of polynomial regression for estimating
free energy differences using thermodynamic integration. We employ linear
regression to construct a polynomial that optimally fits the thermodynamic
integration data, and thus reduces the bias and uncertainty of the resulting
free energy estimate. Two test systems with analytical solutions were used to
verify the accuracy and precision of the approach. Our results suggest that
regression with a high degree of polynomials give the most accurate free energy
difference estimates, but often with a slightly larger variance, compared to
commonly used quadrature techniques. High degrees of polynomials possess the
flexibility to closely fit the thermodynamic integration data but are often
sensitive to small changes in data points. To further improve overall accuracy
and reduce uncertainty, we also examine the use of Chebyshev nodes to guide the
selection of non-equidistant $\lambda$ values for the thermodynamic integration
scheme. We conclude that polynomial regression with non-equidistant $\lambda$
values delivers the most accurate and precise free energy estimates for
thermodynamic integration data. Software and documentation is available at
\url{http://www.phys.uidaho.edu/ytreberg/software}.
\end{abstract}

%
\section{Introduction}
Free energy constitutes an important thermodynamic quantity necessary for a
complete understanding of most chemical and biochemical processes. Examples
such as conformational equilibria and molecular association, partitioning
between immiscible liquids, receptor-drug interaction, protein-protein, and
protein-DNA association, and protein stability all require the underlying free
energy profiles as the prerequisite for a complete comprehension of the
intrinsic properties \cite{chipot07, reddy01, sandal08}. Indeed, the grand
challenge of molecular modeling is to obtain the microscopic detail that is
often inaccessible to conventional experimental techniques. Free energy is
typically expressed as the Helmholtz free energy for an isothermal-isochoric
system or the Gibbs free energy for an isothermal-isobaric system, respectively
\cite{chipot07}.

Thermodynamic integration (TI) has been widely employed to calculate free
energy differences ($\Delta F$) between two well-defined systems
\cite{kirkwood35, mordasini00, shirts05, shirts03, ytreberg06}. It is a general
scheme for the calculation of $\Delta F$ between two systems with potential
energy functions $U_1$ and $U_0$, respectively. The free energy difference,
$\Delta F = F_1 - F_0$, is the reversible work done when the potential energy
function $U_0$ is continuously and reversibly switched to $U_1$, and is defined
as
\begin{equation}
\Delta F = - k_B T \ln \left( \frac{Z \left[ U_1 ( {\vec R} ) \right]}{Z \left[
U_0 ( {\vec R} ) \right]} \right),
\end{equation}
where $k_B$ is the Boltzmann constant, $T$ absolute temperature of the system
in Kelvin, and the configurational partition function is given by
\begin{equation}
Z \left[ U ( {\vec R} ) \right] = \int e^{- U ( {\vec R} ) / k_B T} d {\vec R},
\end{equation}
where ${\vec R}$ is the full set of configuration coordinates. TI is a method
that computes the $\Delta F$ between two systems or states of interest by
estimating the integral
\begin{equation}
\Delta F = \int_{\lambda = 0}^{1} \left \langle \frac{\partial U _{\lambda} (
{\vec R} )}{\partial \lambda} \right \rangle _{\lambda} d \lambda,
\label{equ:integral}
\end{equation}
which is equivalent to the reversible work to switch from $U_0 \to U_1$. The
notation $\left \langle \cdot \right \rangle _{\lambda}$ represents an ensemble
average at a particular value of $\lambda$. Switching the system between two
potential energies requires a continuously variable energy function
$U_{\lambda}$ such that $U_{\lambda = 0} = U_0$ and $U_{\lambda = 1} = U_1$. In
addition, the free energy function $U_{\lambda}$ must be differentiable with
respect to $\lambda$ for $0 \le \lambda \le 1$ \cite{leech01}.

The relationship of eq.\ (\ref{equ:integral}) is exact, but the integral must
be approximated numerically by performing simulation at various discrete values
of $\lambda$. Typically, these discrete $\lambda$ values are used to convert
the integral to a sum (e.g., using quadrature). If the estimates of $\left
\langle \cdot \right \rangle _{\lambda}$ include large fluctuations, then it is
necessary to perform very long simulations in order to calculate the average
value to sufficient statistical accuracy. In addition, the quantity $\left
\langle \cdot \right \rangle_{\lambda}$ may heavily depend on $\lambda$ so that
a large number of simulations at different $\lambda$ values is needed in order
to estimate the integral with sufficient accuracy.

Typically researchers estimate $\Delta F$ with TI utilizing an arithmetic
technique such as the trapezoidal or Simpson's rule. These numerical methods
work well if the curvature of the TI data is small. The trapezoidal rule, for
example, approximates the area under the curve of a given function with a
trapezoid. Thus, $\Delta F$ is approximated by summing the area of the
trapezoids between $\lambda = 0$ and 1. The trapezoidal rule is intrinsically
simple to use and possesses the advantage that the sign of the error of the
approximation can be determined. The trapezoidal rule will overestimate the
integral of a function with a concave-up curve because the trapezoids include
all the area under the curve as well as the extension above it. Similarly, an
underestimate will likely to occur if the function reveals a concave-down curve
because the areas is accounted for under the curve, but not above. However, the
error is difficult to estimate if the curve includes an inflection point.

Importantly, the accuracy of $\Delta F$ using the trapezoidal rule can only
improve by increasing the number of $\left \langle \cdot \right \rangle
_{\lambda}$ even if the $\left \langle \cdot \right \rangle _{\lambda}$ have
sufficiently converged. However, such a large number of long equilibrium
simulations is not always feasible with limited computational resources.

We previously presented the successful application of polynomial and spline
interpolation techniques for $\Delta F$ estimates via TI \cite{shyu08}. These
techniques demonstrate superior accuracy and precision over trapezoidal
quadrature, and give the best estimates of $\Delta F$ without demanding
additional simulations. However, we also noted the inherent weakness and
limitations of the interpolation techniques. The most important weakness is
that high degree of interpolating polynomials suffer from Runge's phenomenon,
i.e., the approximation errors escalate rapidly as the degree of interpolating
polynomial increases. This phenomenon is attributed to the fact that a data
point at or near the middle of the interval gives a large contribution to the
coefficients close to the endpoints. As a consequence, there is a tradeoff
between having a better fit and obtaining a smooth well-behaved fitting
polynomial \cite{faires93, henry81}.

To alleviate these restrictions on polynomial order, we now introduce the
polynomial regression technique for estimating $\Delta F$ using TI. Our goal is
to reduce the bias and uncertainty in the estimates of $\Delta F$ from
evaluation of the integral which is present even for infinite sampling (i.e.,
unbiased $\left \langle \cdot \right \rangle _{\lambda}$). Thus, we implemented
the least squares method to construct the best-fit polynomial model, and used
the Gaussian elimination method with partial pivoting and scaling to calculate
the optimal coefficients for the polynomial. The best-fit polynomial model
interpolates the free energy slope $\frac{dF}{d \lambda} = \left \langle
\frac{\partial U _{\lambda}}{\partial \lambda} \right \rangle _{\lambda}$ as a
function of $\lambda$. Unlike Lagrange and Newton interpolation techniques
\cite{shyu08}, regression permits the degree of polynomial to vary to better
accommodate the curvature of $\frac{dF}{d \lambda}$ \cite{faires93, jeffreys88,
werner84}. Two one-dimensional test systems with analytical solutions were
constructed as the test cases to examine the accuracy and performance of the
regression technique. We also investigated the use of Chebyshev nodes as
non-equidistant $\lambda$ values for TI. The accuracy and precision of free
energy estimates obtained from equidistant and non-equidistant $\lambda$ values
are compared and contrasted. The results from our simulations suggest that
regression, with sufficiently high degree of polynomials, can improve the
accuracy and reduce bias for $\Delta F$ estimates without demanding additional
simulation. Our study further shows that the use of non-equidistant $\lambda$
values further improves the accuracy and reduced uncertainty of the $\Delta F$
estimate over use of equidistant $\lambda$ values.

%
\section{Theory}
The primary objective of this study is to introduce the mathematical and
statistical framework for the analysis of simulation data from TI using
polynomial regression. The objective is to construct a regression model that
best describes the simulation data, namely $\left \langle \frac{\partial U
_{\lambda}}{\partial \lambda} \right \rangle _{\lambda}$, from each TI
simulation at a fixed $\lambda$. The degree of the polynomial for the
regression model is first determined, and best-fit coefficients are
subsequently estimated using the least squares methods. In the context of free
energy estimates using TI, the functional form of the simulation data is
represented by a series of data points $\left \{ \lambda, \frac{dF}{d \lambda}
\right \}$, and the regression model is then constructed through these data
points.

Regression attempts to delineate the relationship between independent (e.g.,
$\lambda$ values) and dependent variables (e.g., $\frac{dF}{d \lambda}$
estimates) by fitting a linear polynomial to the observed data points. The
regression procedures construct a curve that optimally minimizes the errors
between the estimated and observed values. It is important to note that, unlike
Lagrange and Newton polynomial interpolations, regression does not mandate that
$\lambda$ values cover the entire interval between 0 and 1. Instead, the
$\lambda$ values can be chosen anywhere within $\left[ 0, 1 \right]$ because
regression does not attempt to construct a curve that goes through every data
point exactly. In other words, the polynomial model only describes the tendency
and does not delineate the functional form of the data.

Mathematically, polynomial regression is used to fit data points to the
equation $y = \beta_0 + \beta_1 x + \beta_2 x^2 + \ldots + \beta_n x ^n$, where
$\beta _i$ denotes the $i$th coefficient of the polynomial. The degree of
polynomial, $n$, is independent of the number of data points. The higher order
terms in polynomial equation have the greatest effect on the dependent variable
(e.g., $\frac{dF}{d \lambda}$).

We used the least squares method, which is the most widely employed technique
to calculate the best-fit coefficients for the construction of polynomial model
\cite{bjorck96}. It minimizes the sum of the squares of the deviations between
the theoretical curve and the data points from simulations or empirical
observations. A solution is thus obtained without the use of any iterative
procedures. The solution to the construction of polynomial that best represents
the data points is obtained by solving a system of linear equations generated
from the minimization of errors between the true and approximated values. We
utilize Gaussian elimination which is the most commonly used algorithm for
solving systems of linear equations \cite{golub96, rawlings98}. Gaussian
elimination with partial pivoting and scaling, in particular, offers superb
numerical stability, and thus was used for the current study.

%
\subsection{Mathematical Notation}
Regression analysis generally refers to the study of the relationship between
one or several predictors (independent variables we denote as $x$) and the
response (dependent variable we denote as $y$). In the context of free energy
estimates using TI, the simulation data is represented by a series of data
points $\left \{ \lambda, \frac{d F}{d \lambda} \right \} = \{ x, y \}$, and
the polynomial model is constructed through these data points. The following
sections briefly describe the mathematical definitions and properties of the
techniques, the least squares method and Gaussian elimination, for the
construction of a polynomial model that best represents the free energy
profile.

%
\subsection{Least Squares Method}
\begin{figure}[tbp]
\singlespace \centering
\includegraphics[width=4.8in]{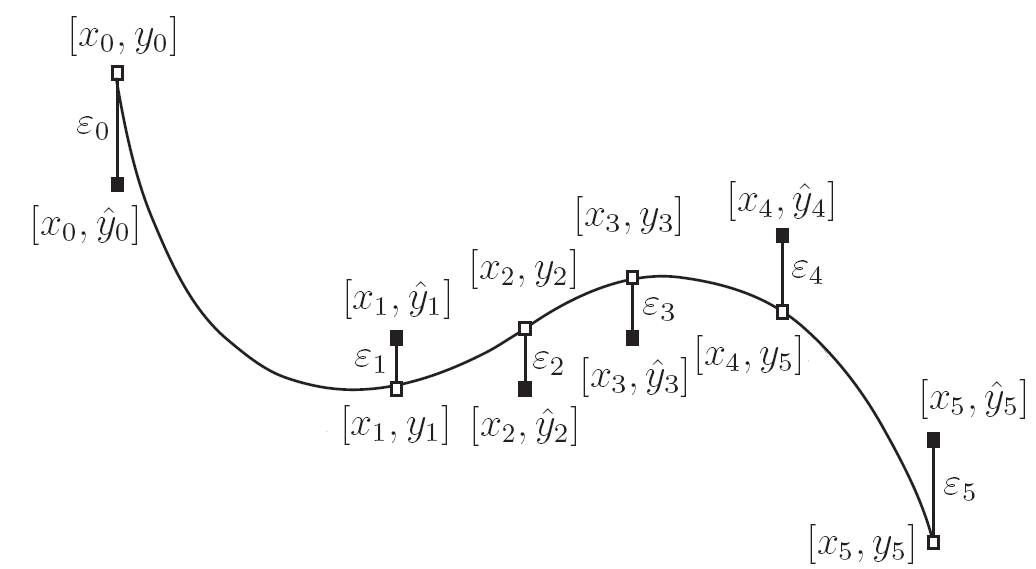} \\
\caption[The least squares approximation technique]{The least squares
approximation technique. The least squares method constructs the best-fit curve
that minimizes the sum of squares of the vertical distances (denoted by
$\varepsilon_0, \varepsilon_1, \ldots, \varepsilon_5$) between the observed
data points $\left[ x_0, \hat y_0 \right], \left[ x_1, \hat y_1 \right],
\ldots, \left[ x_5, \hat y_5 \right]$, and the estimates $\left[ x_0, y_0
\right], \left[ x_1, y_1 \right], \ldots, \left[ x_5, y_5 \right]$ from the
fitting line.} \label{fig:least_squares}
\end{figure}

The least squares method is an approximation technique that constructs the
best-fit curve for a set of data points based on the sum of squares of the
errors. An error at a point is defined as the difference between the true and
approximated values. Geometrically the best-fit curve is the one that minimizes
the sum of squares of the vertical distances between the data points and the
approximating curve (see Fig.\ \ref{fig:least_squares} for an illustration).
The least squares criterion has important statistical interpretations. If
appropriate probabilistic assumptions about underlying error distributions are
made, least squares produce the maximum likelihood estimate of the coefficients
\cite{bjorck96, kopitzke75}. The least squares method has an advantage over the
Lagrange and Newton interpolation techniques \cite{shyu08} as the order of the
approximation is independent of the number of data points. This allows the
degree of polynomial to vary in order to accommodate the desired precision.
Most importantly, least squares does not produce a polynomial that goes through
each data point exactly. The curve that exactly fits all the data points
incorporates all the error in the measurement into the model. Statistically,
this is not a desirable outcome unless the data points have no error, which is
unlikely \cite{rawlings98}.

Approximation using the least squares method can be achieved in either
continuous or discrete forms. We used the discrete least squares method that is
best with applications that have finite data. Specifically it is based on $m$
interpolated data points $\left[ x_i, y_i \right]$ for $i = 0, 1, 2, \ldots,
m$. The curve to be fitted is a polynomial $p \left( x \right)$ that best
represents all data points (e.g., $\lambda$ values).

In the current study, the construction of a polynomial that best represents the
$\lambda$ values was obtained by solving a system of linear equations generated
from the minimization of errors between the true and approximated values.
Mathematically the minimum of the sum of squares was found by setting the
gradient to zero. The process was repeated and applied to all polynomial
coefficients.

For example, consider a least squares approximation using a quadratic
polynomial, $p \left( x \right) = \beta_0 + \beta_1 x + \beta_2 x^2$, where
$\beta _i$ refers to the $i$th coefficient of the polynomial for the regression
model, produce a system of $3 \times 3$ linear equations. The objective
function, $\Psi$, that minimizes the sum of squares of the errors for the
quadratic polynomial is expressed as
\begin{equation}
\Psi = \sum_{i=0}^m \left[ y_i - \left( \beta_0 + \beta_1 x_i + \beta_2 x_i^2
\right) \right] ^2.
\end{equation}
The approximation is obtained by minimizing the sum of squares of the errors
through $\frac{\partial \Psi}{\partial \beta_0} = 0$, $\frac{\partial \Psi}
{\partial \beta_1} = 0$, and $\frac{\partial \Psi}{\partial \beta_2} = 0$. The
first equation is obtained through the steps,
\begin{eqnarray}
\frac{\partial \Psi}{\partial \beta_0} = -2 \sum_{i=0}^m \left[ y_i -
\left( \beta_0 + \beta_1 x_i + \beta_2 x_i^2 \right) \right] = 0, \nonumber \\
\sum_{i=0}^m y_i = \beta_0 \sum_{i=0}^m 1 + \beta_1 \sum_{i=0}^m x_i + \beta_2
\sum_{i=0}^m x_i^2.
\end{eqnarray}
The second equation is generated in the same manner,
\begin{eqnarray}
\frac{\partial \Psi}{\partial \beta_1} = -2 \sum_{i=0}^m x_i \left[ y_i -
\left( \beta_0 + \beta_1 x_i + \beta_2 x_i^2 \right) \right] = 0, \nonumber \\
\sum_{i=0}^m x_i y_i = \beta_0 \sum_{i=0}^m x_i + \beta_1
\sum_{i=0}^m x_i^2 + \beta_2 \sum_{i=0}^m x_i^3.
\end{eqnarray}
Finally, the third equation is obtained from the steps
\begin{eqnarray}
\frac{\partial \Psi}{\partial \beta_2} = -2 \sum_{i=0}^m x_i^2 \left[ y_i
- \left( \beta_0 + \beta_1 x_i + \beta_2 x_i^2 \right) \right] = 0, \nonumber \\
\sum_{i=0}^m x_i^2 y_i = \beta_0 \sum_{i=0}^m x_i^2 + \beta_1
\sum_{i=0}^m x_i^3 + \beta_2 \sum_{i=0}^m x_i^4.
\end{eqnarray}
The three equations can be formulated in the matrix form
\begin{equation}
\left[ {\begin{array}{c c c}
\sum \limits_{i=0}^m 1 & \sum \limits_{i=0}^m x_i &
\sum \limits_{i=0}^m x_i^2 \\
\sum \limits_{i=0}^m x_i & \sum \limits_{i=0}^m x_i^2 &
\sum \limits_{i=0}^m x_i^3 \\
\sum \limits_{i=0}^m x_i^2 & \sum \limits_{i=0}^m x_i^3 &
\sum \limits_{i=0}^m x_i^4 \\
\end{array}} \right]
\left[ {\begin{array}{c}
\beta_0 \\ \beta_1 \\ \beta_2 \\
\end{array}} \right] =
\left[ {\begin{array}{c}
\sum \limits_{i=0}^m y_i \\
\sum \limits_{i=0}^m y_i x_i \\
\sum \limits_{i=0}^m y_i x_i^2 \\
\end{array}} \right].
\end{equation}
These procedures can be generalized to a polynomial of arbitrary degree, $n$.
In the context of free energy estimates using TI, the system of linear
equations is written as
\begin{equation}
\left[ {\begin{array}{c c c c c}
\sum \limits_{i=0}^m 1 & \sum \limits_{i=0}^m \lambda _i & \ldots &
\sum \limits_{i=0}^m \lambda _i^{n-1} & \sum \limits_{i=0}^m \lambda _i^n \\
\sum \limits_{i=0}^m \lambda _i & \sum \limits_{i=0}^m \lambda _i^2 & \ldots &
\sum \limits_{i=0}^m \lambda _i^n & \sum \limits_{i=0}^m \lambda _i^{n+1} \\
\ldots & \ldots & \ldots & \ldots & \ldots \\
\sum \limits_{i=0}^m \lambda _i^n & \sum \limits_{i=0}^m \lambda _i^{n+1} & \ldots &
\sum \limits_{i=0}^m \lambda _i^{2n-2} & \sum \limits_{i=0}^m \lambda _i^{2n-1} \\
\sum \limits_{i=0}^m \lambda _i^{n+1} & \sum \limits_{i=0}^m \lambda _i^{n+2} &
\ldots & \sum \limits_{i=0}^m \lambda _i^{2n-1} &
\sum \limits_{i=0}^m \lambda _i^{2n} \\
\end{array}} \right]
\left[ {\begin{array}{c}
\beta_0 \\ \beta_1 \\ \ldots \\ \beta_{n-1} \\ \beta_n \\
\end{array}} \right] =
\left[ {\begin{array}{c}
\sum \limits_{i=0}^m \left \langle \frac{\partial U}{\partial \lambda} \right
\rangle _{\lambda _i} \\
\sum \limits_{i=0}^m \left \langle \frac{\partial U}{\partial \lambda} \right
\rangle _{\lambda _i} \lambda _i \\
\ldots \\
\sum \limits_{i=0}^m \left \langle \frac{\partial U}{\partial \lambda} \right
\rangle _{\lambda _i} \lambda _i^{n-1} \\
\sum \limits_{i=0}^m \left \langle \frac{\partial U}{\partial \lambda} \right
\rangle _{\lambda _i} \lambda _i^n \\
\end{array}} \right],
\label{equ:linear}
\end{equation}
where $m$ refers to the number of TI simulations. Given the set of linear
equations in eq.\ (\ref{equ:linear}), we now need to solve these equations to
obtain the polynomial coefficients that best fit the TI data. We chose to use
Gaussian elimination with partial pivoting and scaling (see the next section
for more details).

%
\subsection{Gaussian Elimination}
The solution to the system of linear equations has been an important numerical
analysis problem as such a system arises in many different fields of research
\cite{suli03}. It is generally desirable to optimally fit a linear mathematical
model to measurements obtained from simulations in order to obtain a better
insight. The objective is then to extract predictions from the measurements and
to reduce the effect of measurement errors. Numerical methods for solving
linear systems are commonly classified as direct and iterative
\cite{rawlings98}. Linear least squares problems are solved by direct methods
and admit a closed-form solution, in contrast to nonlinear least squares
problems, which have to be solved by an iterative procedure. Direct methods
yield the exact solution, assuming the absence of roundoff or other errors, in
a finite number of elementary arithmetic operations. The fundamental method for
direct solution is Gaussian elimination \cite{suli03}. Gaussian elimination is
an efficient and numerically stable algorithm that utilizes elementary row
operations for the solution of systems of linear equations.

We used Gaussian elimination which is the most commonly employed algorithm to
determine the solutions of a system of linear equations, to find the rank of a
matrix, and to calculate the inverse of an invertible square matrix
\cite{bjorck96, golub96, rawlings98}. The Gauss-Jordan method, the matrix
inverse method, the LU factorization method, and the Thomas algorithm are all
modifications or extensions of the Gaussian elimination method \cite{golub96,
stewart98}. To achieve the optimal numerical stability, we incorporated the
partial pivoting and scaling for solving the system of linear equations.
Partial pivoting and scaling are particular important for free energy estimates
using TI as numerical stability is of primary concern especially for a large
number of simulations.

The elimination process involves normalizing the equation above the element to
be eliminated by the element immediately above the element to be eliminated,
which is called the pivot element, multiplying the normalized equation by the
element to be eliminated, and subtracting the result from the equation
containing the element to be eliminated. This process systematically eliminates
terms below the major diagonal, column by column. This process is continued
until all the coefficients below the major diagonal are eliminated. The
elimination procedure fails immediately if the first pivot element is zero. The
procedure may also fail if any subsequent pivot element is zero. Even though
there may be non-zeros on the major diagonal in the original matrix, the
elimination process may create zeros on the major diagonal. The simple
elimination procedure therefore must be modified to avoid zeros on the major
diagonal. This result can be accomplished by rearranging the equations, by
interchanging equations (rows) or variables (columns), before each elimination
step to put the element of largest magnitude on the diagonal. This process is
referred to as pivoting. Interchanging both rows and columns is called full
pivoting. Full pivoting is quite complex, and thus is rarely used
\cite{golub96, rawlings98}. Interchanging only rows is called partial pivoting.
Pivoting eliminates zeros in the pivot element locations during the elimination
process. Pivoting also reduces roundoff errors, since the pivot element is a
divisor during the elimination process, and division by large numbers
introduces smaller roundoff errors than division by small numbers
\cite{golub96}. However, when the procedure is repeated, roundoff errors can
still compound. This problem escalates rapidly as the number of equations
increases.

The elimination process can indeed incur significant roundoff errors when the
magnitudes of the pivot elements are smaller than the magnitudes of the other
elements in the equations containing the pivot elements. In such cases, scaling
is generally employed to select the pivot elements. The process of scaling
involves normalizing all the elements in the first column by the largest
element in the corresponding rows. Pivoting is then implemented based on the
scaled elements in the first column, and elimination is applied to obtain zero
elements in the first column below the pivot element. Similarly, before
elimination is applied to the second column, all of the elements from 2 to $n$
are scaled, pivoting is implemented, and elimination is applied to obtain zero
elements in column 2 below the pivot element. The procedure is applied to the
remaining rows 3 to $n - 1$. Back substitution is then applied to obtain the
solutions. The effects of roundoff can be reduced by scaling the equations
before pivoting. It is worth noting that scaling itself sometimes can introduce
additional roundoff errors and should be used only to determine if pivoting is
required. Scaling is generally not required if all the elements of the
coefficient matrix are the same order of magnitude. For optimal numerical
stability, we implemented the scaling algorithm to avoid pivoting to zero pivot
elements \cite{golub96, stewart73}.

%
\subsection{Chebyshev Nodes for Non-equidistant $\lambda$ Value Selection}
Curve fitting with high degree of polynomials has not been a popular subject
because such curves are particulary sensitive to small changes in the
coefficients. Studies have shown that polynomial regression using equidistant
abscissas, in particular, can give rise to convergence difficulties, especially
with high degree of polynomials \cite{faires93, werner84}. A data point at or
near the middle of the interval gives a large contribution to the values of $p
\left( x \right)$ close to the endpoints. In other words, a small change to a
data point in the middle can produce a significant excursion in the curve near
the ends. The phenomenon is problematic with a set of a dozen or more data
points that are more or less equidistant along the interval. Regression models
with high order terms (typically higher than four) become more sensitive to the
precision of coefficient values, where small differences in the coefficient
values can result in a large differences in the computed $y$ value
\cite{faires93}. This difficulty intimates that high degree polynomials can be
very sensitive to disturbances in the given values of the function.

Mathematically, equidistant curve fitting using polynomials of high degree is
in some causes an ill-conditioned problem, especially in the outer parts of the
interval $\left[ x_0, x_m \right]$ \cite{crouse64}. An ill-conditioned problem
is one in which a small change in any of the elements of the problem causes a
large fluctuation in the solution of the problem. For polynomial regression
using the least squares method, for example, the coefficients in the matrix can
vary over a wild range of several orders of magnitude. Since ill-conditioned
systems are sensitive to small changes in the elements of the problem, they are
also sensitive to roundoff errors. Werner \cite{werner84} reported that special
arrangements on the abscissae must be made in order to avoid the fluctuations
on high degree of polynomials. Chebyshev nodes have been widely employed to
counter the effects of such fluctuations \cite{suli03}. Mathematically,
Chebyshev nodes are derived from the roots of Chebyshev polynomial of the first
kind, and tend to concentrate more heavily at the beginning and end of the
interval. The use of Chebyshev nodes guarantees that the maximum error
diminishes as the degree of polynomial increases \cite{faires93}.

\begin{figure}[tb]
\singlespace \centering
\includegraphics[width=4.8in]{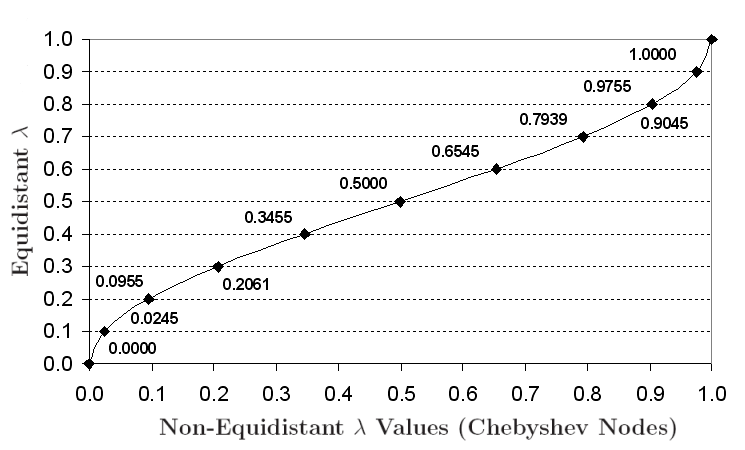} \\
\caption[Transformation from equidistant to non-equidistant $\lambda$ values]{
Transformation from equidistant to non-equidistant $\lambda$ values. The use of
Chebyshev nodes for guiding the selection of non-equidistant $\lambda$ values
aims to reduce the numerical instability of the regression as the degree of
interpolating polynomial increases. For TI simulations, the non-equidistant
$\lambda$ values are selected based on the expression $\lambda = \frac{1}{2}
\left| \cos \left( \varphi \pi \right) - 1 \right|$.} \label{fig:chebyshev}
\end{figure}

For TI simulations, Chebyshev nodes (non-equidistant $\lambda$ values) are
chosen using the following expression
\begin{equation}
\lambda = \frac{1}{2} \left| \cos \left( \varphi \pi \right) - 1 \right|,
\end{equation}
for $\varphi \in \left[ 0, 1 \right]$. Chebyshev nodes possess the property
that they become close together near the boundaries of the region (see Fig.\
\ref{fig:chebyshev} for an illustration). The use of Chebyshev nodes to guide
the selection of non-equidistant $\lambda$ values aims to stabilize the
polynomial construction processes. This is particularly important when the
regression model includes a high degree of polynomial.

We note that several studies have reported the use of non-equidistant $\lambda$
values to improve the overall accuracy and precision of free energy estimate
\cite{blondel04, koning05, mark90, schmiedl07, shobana00, steinbrecher07}.
However, it is important to realize that these studies used non-equidistant
$\lambda$ values to better describe the curvature of the energy slope, while we
are using non-equidistant $\lambda$ values to improve the numerical stability
and accuracy of the construction of polynomials that fit the TI data.

%
%
\subsection{Statistical Implications of Polynomial Modeling}
Approximation of complex functions by polynomials is indeed a basic building
block for a great many numerical techniques \cite{suli03}. The degree of
realism that needs to be incorporated into a model largely depends on the
purpose of the regression analysis. The \emph{F}-static is commonly employed to
test the null hypothesis such as, $H_0 : \beta_{q + 1} = \beta_{q + 2} = \ldots
= \beta_{m - 1} = 0$, for $0 \le q \le m$, for the adequacy of a model
\cite{rawlings98}. The least demanding purpose is the simple use of a
regression model to summarize the observed relationship in a particular set of
data points \cite{bjorck96, chambers73}. There is no interest in the functional
form of the model or in predictions to other sets of data or situations. The
paramount concern is that the model adequately portrays the observed
relationships. TI simulations data fall into this category.

It is very unlikely that the free energy changes diligently follows a
particular functional form. However, it is crucial that the polynomial model
closely fits the ensemble averages obtained from each TI simulation. The most
demanding task using regression, on the other hand, is the esoteric development
of mathematical models to accurately describe the physical, chemical, and
biological processes in the system. The objective is to make the model as
realistic as the state of knowledge will permit. Realistic models will tend to
provide more protection against large errors from experiments or simulations.
In the context of TI simulations, however, realistic models will likely mandate
infinitely long simulations. Some authors \cite{rawlings98, crouse64} reported
that realistic models often may be simpler in terms of the number of
coefficients to be estimated. A response with a plateau, for example, may
require several terms of a polynomial to fit the plateau, but might be
characterized very well with a two-coefficient exponential model.

%
\section{Computational Details}
\begin{figure}[tbp]
\singlespace \centering
\includegraphics[width=5.0in]{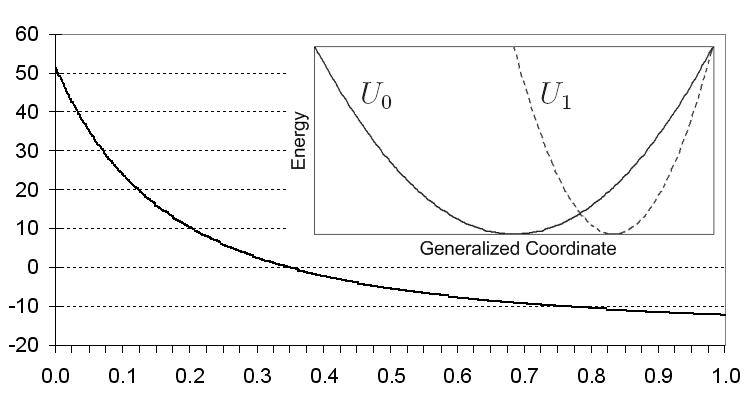} \\
\caption[Free energy slope, $\frac{dF}{d \lambda} = \left \langle
\frac{\partial U _{\lambda}}{\partial \lambda} \right \rangle _{\lambda}$, of
the first one-dimensional test system obtained from TI simulations using 1001
equidistant $\lambda$ values]{Free energy slope, $\frac{dF}{d \lambda} = \left
\langle \frac{\partial U _{\lambda}}{\partial \lambda} \right \rangle
_{\lambda}$, of the first one-dimensional test system obtained from TI
simulations using 1001 equidistant $\lambda$ values. The figure on the top
right corner shows that the potential energy functions, $U_0 \left( \xi \right)
= \frac{1}{2} \xi ^2$ (solid line) and $U_1 \left( \xi \right) = 2 \left( \xi -
5 \right) ^2$ (dashed line), generate two offset harmonic wells with different
curvature. The analytical solution of this system is $- \frac{1}{2} \ln
\frac{1}{4}$.} \label{fig:skewed}
\end{figure}

\begin{figure}[tbp]
\singlespace \centering
\includegraphics[width=5.0in]{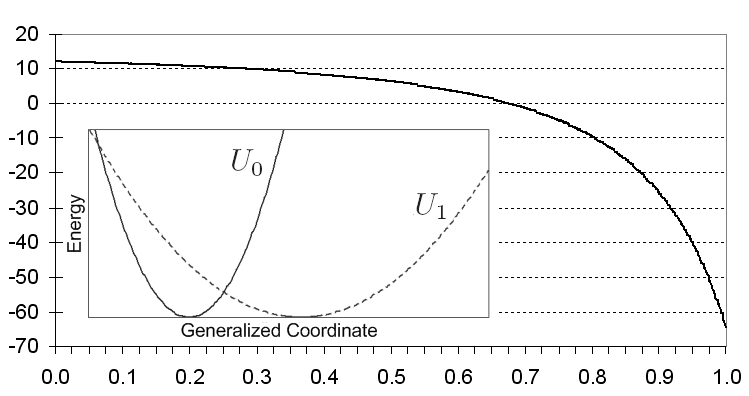} \\
\caption[Free energy slope, $\frac{dF}{d \lambda} = \left \langle
\frac{\partial U _{\lambda}}{\partial \lambda} \right \rangle _{\lambda}$, of
the second one-dimensional test system obtained from TI simulations using 1001
equidistant $\lambda$ values]{Free energy slope, $\frac{dF}{d \lambda} = \left
\langle \frac{\partial U _{\lambda}}{\partial \lambda} \right \rangle
_{\lambda}$, of the second one-dimensional test system obtained from TI
simulations using 1001 equidistant $\lambda$ values. The figure on the bottom
left corner shows that the potential energy functions, $U_0 \left( \xi \right)
= \frac{5}{2} \xi ^2$ (solid line) and $U_1 \left( \xi \right) = \frac{1}{2}
\left( \xi - 5 \right) ^2$ (dashed line), generate two offset harmonic wells
with different curvature. The analytical solution of this system is $-
\frac{1}{2} \ln 5$.} \label{fig:steep}
\end{figure}

Two test systems were constructed to analyze the accuracy and precision of the
regression techniques for estimating $\Delta F$ using TI data. For the purpose
of this study, it is important to use systems with an analytical solution in
order to provide an objective analysis of the accuracy and precision of
regression techniques. The first system involves two potential functions $U_0
\left( \xi \right) = \frac{1}{2} \xi ^2$ and $U_1 \left( \xi \right) = 2 \left(
\xi - 5 \right) ^2$, where $\xi$ is the position of the particle, (see Fig.\
\ref{fig:skewed} for an illustration of the free energy curve) and the second
system uses $U_0 \left( \xi \right) = \frac{5}{2} \xi^2$ and $U_1 \left( \xi
\right) = \frac{1}{2} \left( \xi - 5 \right) ^2$ (see Fig.\ \ref{fig:steep} for
an illustration of the free energy curve). The slope of $\frac{dF}{d \lambda}$
is considerably steeper for the second system and thus a much larger number of
TI simulations would be required in order to achieve accuracy similar to that
of the first system when using quadrature.

For our simulations, the switching function was defined as $U_{\lambda} \left(
\xi \right) = \lambda U_1 \left( \xi \right) + \left( 1 - \lambda \right) U_0
\left( \xi \right)$. The non-equidistant $\lambda$ values are chosen according
to the expression $\lambda = \frac{1}{2} \left| \cos \left( \varphi \pi \right)
- 1 \right|$. The analytical solution of the $\Delta F$ for the first system is
$- \frac{1}{2} \ln \frac{1}{4}$, and second $- \frac{1}{2} \ln 5$.

Simulations were performed with two sets of equidistant and non-equidistant
$\lambda$ values, and TI simulations were performed at each value of $\lambda$.
A total of 1,000 independent trials was run for each system. Simulations were
performed for six and eleven equidistant and non-equidistant $\lambda$ values.
An equal amount of simulation times (1,000,000 Monte Carlo steps) were used for
each of six ($\lambda = 0.0, 0.2, 0.4, 0.6, 0.8$, and $1.0$) and eleven
($\lambda = 0.0, 0.1, 0.2, \ldots, 0.9$, and $1.0$) equidistant values of
$\lambda$. Identical TI simulations were performed on the corresponding
non-equidistant $\lambda$ values, for six ($\lambda$ = 0.0, 0.0955, 0.3455,
0.6545, 0.9045, and 1.0) and eleven ($\lambda$ = 0.0, 0.0245, 0.0955, 0.2061,
0.3455, 0.5, 0.6545, 0.7939, 0.9045, 0.9755, and 1.0). Averages of the slope
$\frac{dF}{d \lambda} = \left \langle \frac{\partial U _{\lambda}}{\partial
\lambda} \right \rangle _{\lambda}$ were collected for each value of $\lambda$.

Each simulation started with an arbitrarily chosen initial position for the
particle. Metropolis Monte Carlo \cite{metropolis53} was performed using trial
moves generated by adding a uniform random deviate between -0.5 and 0.5 to the
current position. The acceptance ratio was maintained in the range of
approximately 40 to 45\% for each trial. Simulations were given 1,000 steps to
equilibrate initially, then were allowed to proceed until the desired number of
Monte Carlo steps (1,000,000) has been reached.

%
\section{Results and Discussion}
Table \ref{fig:s1l05_table} summarizes averages and standard deviations of the
biases for the estimates of $\Delta F$ from 1,000 independent trials on the
first test system (see Fig.\ \ref{fig:skewed} for an illustration of the energy
curve) using six equidistant and non-equidistant $\lambda$ values. We defined
the bias as the difference between the analytical and estimated free energy.
The bias from each independent trial was recorded, and averages and standard
deviations were calculated for comparison. For higher degrees of polynomials,
the estimates obtained from equidistant $\lambda$ values begin to diverge
markedly. The averages and standard deviations of the biases fluctuate widely.
This is largely due to the effects of overfitting. Overfitting occurs when a
model contains too many parameters. An unrealistic model may fit perfectly if
the model has enough complexity by comparison to the amount of data available.
In other words, when the degrees of freedom in parameter selection exceed the
information content of the data, this leads to arbitrariness in the final model
parameter which reduces or destroys the ability of the model to generalize
beyond the fitting data. The likelihood of overfitting depends not only on the
number of parameters and data but also the conformability of the model
structure with the data shape \cite{bjorck96, kopitzke75, davis33, lewis47}.
This phenomenon, however, does not appear on the estimates obtained from
simulations using the non-equidistant $\lambda$ values. Overall, the estimates
of $\Delta F$ obtained from simulations using non-equidistant $\lambda$ values
are considerably more accurate than that of equidistant $\lambda$ values. The
estimates obtained from polynomials of degree six, for example, using
non-equidistant $\lambda$ values are more accurate than that of degree ten
using equidistant $\lambda$ values. It is worth noting that although the
estimates obtained from polynomials of degree seven and higher show superior
accuracy, these estimates, however, are no longer reliable as the number of
parameters in the regression modes exceeds the number of $\lambda$ values.

\begin{table}[tbp]
\singlespace \centering
\begin{tabular}{|r|r r|r r|}
\multicolumn{5}{c}{Biases for $\Delta F$ estimates using six $\lambda$ values} \\
\multicolumn{5}{c}{for the first test system} \\
\hline
& \multicolumn{2}{c|}{Equidistant $\lambda$} &
\multicolumn{2}{c|}{Non-equidistant $\lambda$} \\
 & Average & Std. Dev. & Average & Std. Dev. \\
\hline
Trapezoid & 1.2490 & 0.0203 & 1.1398 & 0.0209 \\
Degree 3 & 0.6893 & 0.0193 & 0.2520 & 0.0196 \\
Degree 4 & 0.1599 & 0.0196 & -0.0201 & 0.0211 \\
Degree 5 & 0.1599 & 0.0196 & -0.0201 & 0.0211 \\
Degree 6 & -0.6105 & 0.0426 & -0.0296 & 0.0212 \\
Degree 7 & 0.2537 & 0.0239 & -0.0430 & 0.0213 \\
Degree 8 & 2.3583 & 0.1555 & -0.0150 & 0.0213 \\
Degree 9 & 0.6152 & 0.0224 & 0.0032 & 0.0214 \\
Degree 10 & -3.4197 & 0.2292 & -0.4076 & 0.0269 \\
\hline
\end{tabular}
\caption[Averages and standard deviation of the biases for the estimates of
$\Delta F$ of the first test system (see Fig.\ \ref{fig:skewed} for an
illustration of the energy curve) using six equidistant and non-equidistant
$\lambda$ values]{Averages and standard deviation of the biases for the
estimates of $\Delta F$ of the first test system (see Fig.\ \ref{fig:skewed}
for an illustration of the energy curve) using six equidistant and
non-equidistant $\lambda$ values. The degrees of polynomials vary from three to
ten.} \label{fig:s1l05_table}
\end{table}

\begin{table}[tbp]
\singlespace \centering
\begin{tabular}{|r|r r|r r|}
\multicolumn{5}{c}{Biases for $\Delta F$ estimates using eleven $\lambda$ values} \\
\multicolumn{5}{c}{for the first test system} \\
\hline
& \multicolumn{2}{c|}{Equidistant $\lambda$} &
\multicolumn{2}{c|}{Non-equidistant $\lambda$} \\
 & Average & Std. Dev. & Average & Std. Dev. \\
\hline
Trapezoid & 0.3268 & 0.0145 & 0.3078 & 0.0154 \\
Degree 3 & 0.3757 & 0.0144 & 0.2226 & 0.0146 \\
Degree 4 & 0.0562 & 0.0143 & 0.0167 & 0.0152 \\
Degree 5 & 0.0562 & 0.0143 & 0.0167 & 0.0152 \\
Degree 6 & 0.0121 & 0.0150 & 0.0025 & 0.0154 \\
Degree 7 & 0.0121 & 0.0150 & 0.0025 & 0.0154 \\
Degree 8 & 0.0044 & 0.0166 & 0.0013 & 0.0155 \\
Degree 9 & 0.0044 & 0.0166 & 0.0013 & 0.0155 \\
Degree 10 & 0.0033 & 0.0347 & 0.0012 & 0.0155 \\
\hline
\end{tabular}
\caption[Averages and standard deviation of the biases for the estimates of
$\Delta F$ of the first test system (see Fig.\ \ref{fig:skewed} for an
illustration of the energy curve) using eleven equidistant and non-equidistant
$\lambda$ values]{Averages and standard deviation of the biases for the
estimates of $\Delta F$ of the first test system (see Fig.\ \ref{fig:skewed}
for an illustration of the energy curve) using eleven equidistant and
non-equidistant $\lambda$ values. The degrees of polynomials vary from three to
ten.} \label{fig:s1l10_table}
\end{table}

Table \ref{fig:s1l10_table} summarizes averages and standard deviations of the
biases for the estimates of $\Delta F$ from 1,000 independent trials for the
first test system (see Fig.\ \ref{fig:skewed} for an illustration of the energy
curve) using eleven equidistant and non-equidistant $\lambda$ values. Overall,
polynomial regression has made a significant improvement on the estimates using
eleven $\lambda$ values. The results show that polynomials of degree eight seem
sufficient to accurately estimate $\Delta F$. The estimates obtained from
polynomials of higher degrees using equidistant $\lambda$ values, however,
include much larger variation. As the degrees of polynomials increase, the
standard deviations also increase. Using equidistant $\lambda$ values, higher
degrees of polynomials are more likely to produce estimates with large
fluctuations which inevitably degrade the overall accuracy and precision. This
phenomenon reflects the fact that high degree of polynomials are sensitive to
small changes in the data. Polynomial regression using non-equidistant
$\lambda$ values, on the other hand, revels superior overall performance on the
estimates of $\Delta F$. The estimates derived from polynomials using
non-equidistant $\lambda$ values are much more accurate, and include much
smaller variations. The standard deviations of the biases remain relatively
constant. The overall estimates are much more accurate than that of equidistant
$\lambda$ values. As an example, the estimates obtained from polynomials of
degree six using non-equidistant $\lambda$ values are more accurate than that
of degree ten using equidistant $\lambda$ values.

Table \ref{fig:s2l05_table} summarizes averages and standard deviations of the
biases for the estimates of $\Delta F$ from 1,000 independent trials using six
equidistant and non-equidistant $\lambda$ values for the second test system
(see Fig.\ \ref{fig:steep} for an illustration of the energy curve). The
estimates obtained from polynomials using equidistant $\lambda$ values are all
heavily biased. For polynomials of higher degrees, the estimates become
unreliable and include substantial variations. Similar to that of the first
test system, overfitting causes the estimates of $\Delta F$ to fluctuate
profoundly. For the estimates obtained from simulations using the
non-equidistant $\lambda$ values, the estimates are still somewhat biased but
greatly reduced from other estimates. As the degree of polynomials increases,
the accuracy also improves accordingly. Since the second test system bears a
much steeper energy curve, the variances of estimates are slightly larger than
that of the first test system. It is clear that the use of non-equidistant
$\lambda$ values gives much accurate estimates of $\Delta F$. This is
apparently due to the fact that the use of non-equidistant $\lambda$ values
significantly stabilize the polynomial construction processes and,
subsequently, improve the overall accuracy.

\begin{table}[tbp]
\singlespace \centering
\begin{tabular}{|r|r r|r r|}
\multicolumn{5}{c}{Biases for $\Delta F$ estimates using six $\lambda$ values} \\
\multicolumn{5}{c}{for the second test system} \\
\hline
& \multicolumn{2}{c|}{Equidistant $\lambda$} &
\multicolumn{2}{c|}{Non-equidistant $\lambda$} \\
 & Average & Std. Dev. & Average & Std. Dev. \\
\hline
Trapezoid & -1.8947 & 0.0236 & -1.4949 & 0.0223 \\
Degree 3 & -1.2068 & 0.0221 & -0.4202 & 0.0205 \\
Degree 4 & -0.3495 & 0.0212 & 0.0467 & 0.0218 \\
Degree 5 & -0.3495 & 0.0212 & 0.0467 & 0.0218 \\
Degree 6 & -0.7456 & 0.0623 & 0.0386 & 0.0218 \\
degree 7 & -0.2243 & 0.0363 & 0.0351 & 0.0218 \\
Degree 8 & 0.2189 & 0.3169 & 0.0278 & 0.0219 \\
Degree 9 & -0.2006 & 0.0324 & 0.0260 & 0.0222 \\
Degree 10 & 0.0571 & 0.5375 & 0.0306 & 0.0340 \\
\hline
\end{tabular}
\caption[Averages and standard deviation of the biases for the estimates of
$\Delta F$ of the second test system (see Fig.\ \ref{fig:steep} for an
illustration of the energy curve) using six equidistant and non-equidistant
$\lambda$ values]{Averages and standard deviation of the biases for the
estimates of $\Delta F$ of the second test system (see Fig.\ \ref{fig:steep}
for an illustration of the energy curve) using six equidistant and
non-equidistant $\lambda$ values. The degrees of polynomials vary from three to
ten.} \label{fig:s2l05_table}
\end{table}

\begin{table}[tbp]
\singlespace \centering
\begin{tabular}{|r|r r|r r|}
\multicolumn{5}{c}{Biases for $\Delta F$ estimates using eleven $\lambda$ values} \\
\multicolumn{5}{c}{for the second test system} \\
\hline
& \multicolumn{2}{c|}{Equidistant $\lambda$} &
\multicolumn{2}{c|}{Non-equidistant $\lambda$} \\
 & Average & Std. Dev. & Average & Std. Dev. \\
\hline
Trapezoid & -0.5076 & 0.0164 & -0.4072 & 0.0158 \\
Degree 3 & -0.6552 & 0.0167 & -0.3754 & 0.0150 \\
Degree 4 & -0.1257 & 0.0158 & -0.0331 & 0.0154 \\
Degree 5 & -0.1257 & 0.0158 & -0.0331 & 0.0154 \\
Degree 6 & -0.0325 & 0.0164 & -0.0022 & 0.0157 \\
Degree 7 & -0.0325 & 0.0164 & -0.0022 & 0.0157 \\
Degree 8 & -0.0116 & 0.0178 & 0.0012 & 0.0158 \\
Degree 9 & -0.0116 & 0.0178 & 0.0012 & 0.0158 \\
Degree 10 & -0.0051 & 0.0332 & 0.0016 & 0.0159 \\
\hline
\end{tabular}
\caption[Averages and standard deviation of the biases for the estimates of
$\Delta F$ of the second test system (see Fig.\ \ref{fig:steep} for an
illustration of the energy curve) using eleven equidistant and non-equidistant
$\lambda$ values]{Averages and standard deviation of the biases for the
estimates of $\Delta F$ of the second test system (see Fig.\ \ref{fig:steep}
for an illustration of the energy curve) using eleven equidistant and
non-equidistant $\lambda$ values. The degrees of polynomials vary from three to
ten.} \label{fig:s2l10_table}
\end{table}

Table \ref{fig:s2l10_table} summarizes the averages of biases for the estimates
of $\Delta F$ obtained from 1,000 independent trials using eleven equidistant
and non-equidistant $\lambda$ values for the second test system (see Fig.\
\ref{fig:steep} for an illustration of the energy curve). With eleven rather
than six $\lambda$ values, polynomials of degree eight seem sufficient to
accurately estimate $\Delta F$ using either equidistant or non-equidistant
$\lambda$ values. The estimates obtained from simulations using non-equidistant
$\lambda$ values are generally more accurate than that of equidistant $\lambda$
values. It is worth noting that the estimates obtained from simulations using
equidistant $\lambda$ values generally include much larger variations. This
reflects that fact that high degrees of polynomials are more susceptible to
oscillation. This is particular evident when the energy profile bears a steep
curvature. Special arrangement for the $\lambda$ values should be made if one
wishes to reduce the oscillation and improve the accuracy of approximation.
Polynomial regression using non-equidistant $\lambda$ values shows superior
overall performance. The estimates are considerably more accurate than that of
equidistant $\lambda$ values. The use of non-equidistant $\lambda$ values
reveals strong potential for the estimates of $\Delta F$ using polynomial
regression. The estimates obtained from the trapezoidal quadrature using either
the equidistant or non-equidistant $\lambda$ values, however, are still heavily
biased.

%
\section{Conclusion}
We utilized polynomial regression to estimate free energy differences from
thermodynamic integration simulation data. Two test systems were used to
validate the accuracy and precision of the regression technique. The least
squares method was used to extract vital information from each thermodynamic
integration simulation and construct globally optimal polynomial model that
best fits the free energy profile with respect to the switching variable
$\lambda$. Gaussian elimination with partial pivoting and scaling was
implemented to solve the resulting system of linear equations.

Results from the simulations clearly show that the use of regression with high
degrees of polynomials gives the most accurate estimates of free energy
differences. Although it is unlikely that the simulation data from
thermodynamic integration will possess true polynomial representation, the
flexibility of high degree polynomials allows the regression model to be
approximated to any desired precision. Regression possesses the unique
advantage that it permits the degree of polynomial to vary independently of the
number of $\lambda$ values. This property significantly alleviate the
restriction imposed by polynomial interpolation techniques such as Lagrange and
Newton polynomial used in a previous study \cite{shyu08}. Estimates of $\Delta
F$ for a large number of $\lambda$ values is there fore more straightforward.
However, we caution that the degree of polynomial in regression model should be
chosen to limit to a dozen or less to ensure numerical stability.

We also investigated the use of Chebyshev nodes (non-equidistant $\lambda$
values) for $\Delta F$ estimates and found that it improves the overall
accuracy and reduces the biases compared to that of equidistant $\lambda$
values. Our results clearly demonstrate that the use of non-equidistant
$\lambda$ values significantly reduces the variance, and improves the overall
accuracy of the free energy estimates compared to that of equidistant $\lambda$
values. Our study has confirmed that the use of Chebyshev nodes to guide the
choice of $\lambda$ values, in particular, offer superior numerical stability
for regression analysis of TI data. Thus researchers are encouraged to use the
polynomial regression and non-equidistant $\lambda$ values for their future
free energy simulation using thermodynamic integration.

To allow researchers to immediately utilize the method, free software and
documentation is provided via
\url{http://www.phys.uidaho.edu/ytreberg/software}.

%
\section*{Acknowledgements}
This research was supported by the University of Idaho NSF-EPSCoR, and
Bionanoscience at UI (BANTech).

%
\bibliographystyle{plain}
\singlespace \bibliography{article}

\end{document}